\documentclass[
12pt,
floatfix,
amssymb,
pra
]{iopart}

\usepackage{iopams}
\usepackage{url}
\expandafter\let\csname equation*\endcsname\relax
\expandafter\let\csname endequation*\endcsname\relax
\usepackage{amsmath}
\usepackage{graphicx}
\usepackage{upgreek}
\usepackage{appendix}
\usepackage{multirow}
\usepackage{tabu}
\usepackage{float}
\usepackage{xcolor}

\begin{document}

\newcommand{\red}{\textcolor{red}}
\title{Cryogenic electro-optic modulation in titanium in-diffused lithium niobate waveguides}

\author{Frederik Thiele$^1$, Felix vom Bruch $^2$, Julian Brockmeier $^1$, Maximilian Protte $^1$, Thomas Hummel $^1$, Raimund Ricken $^2$, Victor Quiring $^2$, Sebastian Lengeling $^2$, Harald Herrmann $^2$, Christof Eigner $^2$, Christine Silberhorn $^2$, and Tim J. Bartley $^1$}
\address{$^1$ Mesoscopic Quantum Optics, Department Physics, Paderborn  University, 33098 Paderborn Warburger Str. 100, Germany
}%
\address{ 
	$^2$ Integrated Quantum Optics, Department Physics, Paderborn  University, 33098 Paderborn Warburger Str. 100, Germany
}%

\begin{abstract}
    Lithium niobate is a promising platform for integrated quantum optics. In this platform we aim to efficiently manipulate and detect quantum states by combining superconducting single photon detectors and modulators. The cryogenic operation of a superconducting single photon detector dictates the optimisation of the  electro-optic modulators under the same operating conditions. To that end, we characterise a phase modulator, directional coupler, and polarisation converter at both ambient and cryogenic temperatures. The operation voltage $V_{\pi/2}$ of these modulators increases due to the decrease of the electro-optic effect by 74\% for the phase modulator, 84\% for the directional coupler and 35\% for the polarisation converter below 8.5$\,\mathrm{K}$. The phase modulator preserves its broadband nature and modulates light in the characterised wavelength range. The unbiased bar state of the directional coupler changed by a wavelength shift of 85$\,\mathrm{nm}$ while cooling the device down to 5$\,\mathrm{K}$. The polarisation converter uses periodic poling to phasematch the two orthogonal polarisations. The phasematched wavelength of the used poling changes by 112$\,\mathrm{nm}$ when cooling to 5$\,\mathrm{K}$.
    	
\end{abstract}

\section{Introduction}
    Quantum photonics is a promising field since it enables and improves photonic technologies such as secure quantum communication \cite{Gisin2007}, quantum computation \cite{Zhong2020}, and metrology \cite{Polino2020,You2021}. Significant improvements can be made by integrating multiple photonic elements in a single unit. Integration reduces the footprint, increases circuit complexity, and minimises interface losses and power consumption, thus enhancing scalability~\cite{Bogdanov2017,Wang2019a,Elshaari2020,Kim2020,Zhang2021,Moody2021}.
    
    Quantum photonics requires the generation, manipulation, and detection of single photons. Scalable integrated quantum optics requires low system loss, nonlinear effects for photon generation and modulation, as well as efficient single-photon detection. Superconducting Nanowire Single Photon Detectors (SNSPDs) are one of the highest performing single photon detectors~\cite{EsmaeilZadeh2021}, since they achieve a detection efficiency of above 98\%~\cite{Reddy2020}, a low timing jitter~\cite{Korzh2020}, and low dark count rates~\cite{Hochberg2019}. These superconducting detectors require cryogenic operating temperatures and have been integrated in photonic platforms \cite{Ferrari2018}.
    
    An optimal photonic platform needs to combine the integrability of modulators and single photon detectors at cryogenic temperatures \cite{Elshaari2020,DeCea2020,Kim2020,Lomonte2021}. Titanium in-diffused lithium niobate is an interesting platform to pursue this goal since it offers a large second order nonlinearity, low propagation loss, butt-coupling with single mode fibres, and integrability of SNSPDs \cite{Alibart2016b,Sharapova2017a}. To generate and manipulate quantum states, spontaneous parametric down-conversion~\cite{Lange2021} as well as frequency conversion~\cite{Bartnick2021} has recently been shown at cryogenic temperatures in this platform. Furthermore, superconducting single photon detectors such as transition edge sensors \cite{Hopker2019} and SNSPDs have been integrated in lithium niobate \cite{Tanner2012,Smirnov2018,Hopker2020,Lomonte2020,Sayem2020b,Colangelo2020,Lomonte2021}. Finally, lithium niobate offers a large electro-optic effect for optical modulation, which is also functional at cryogenic temperatures \cite{YOSHIDA1992,Morse1994b,Herzog2008,Thiele2020b,Youssefi2021a}.
    
    \begin{figure}[H]
		\centering
		\includegraphics[width=0.7\linewidth]{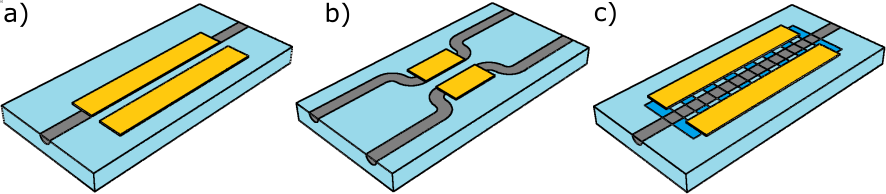}
		\caption{Electro-optic modulators in z-cut lithium niobate waveguides. a) polarisation converter b) directional coupler c) polarisation converter}
		\label{fig:TrioMods}
	\end{figure}
   
    Nevertheless, cryogenic operation of modulators in lithium niobate is a understudied research area. Morse et al. and Yoshida et.al. \cite{YOSHIDA1992,Morse1994b}, independently demonstrated a directional coupler in lithium niobate at low temperature for the first time. The low-temperature electro-optic properties of bulk lithium niobate have been investigated in Ref. ~\cite{Herzog2008}. A cryogenic phase modulator was used in Refs.~\cite{Youssefi2021a,Lomonte2021} to read out a superconducting device. Polarisation conversion under cryogenic conditions was recently demonstrated by our group~\cite{Thiele2020b}. Building on these initial results, in this paper we investigate the performance metrics of three classes of electro-optic modulators in lithium niobate under cryogenic conditions. Specifically, we fabricate and compare a phase modulator, directional coupler, and polarisation converter below 8.5$\,\mathrm{K}$. 
    
    These devices are shown schematically in Fig.~\ref{fig:TrioMods}. Each device functions by coupling an externally applied electric field to modify the optical field(s) propagating in the waveguide. In Sec.~\ref{sec:theory}, we provide a general theory for the operating principle of these electro-optic devices, and their modifications under cryogenic operation. In Sec.~\ref{sec:Fab} we give details on the fabrication of our devices, followed by the general characterisations setup in Sec.~\ref{sec:Char}. The results for each modulator are given in Sec.~\ref{sec:Modulators}, before begin compared in Sec.~\ref{sec:Comp}. Finally, we conclude in Sec.~\ref{sec:concl}.

\section{Theory}\label{sec:theory}
    Optical modulation requires a controlled change of the refractive index. Several effects enable the change of the refractive index for example the electro-optic and acousto-optic effects \cite{Nye1957,Yariv1973,Alferness1982}. To integrate modulators with SNSPDs, the modulators must not only function at cryogenic temperatures, but also be compatible with the cooling power of the cryostat. Acousto-optics \cite{McKenna2020}, thermo-optics \cite{Komma2012b,Harris2014}, carrier injection \cite{Gehl2017}, DC-Kerr \cite{Chakraborty2020} and the electro-optic effect \cite{YOSHIDA1992,Morse1994b,Eltes2020,Thiele2020b,Lee2020} have been shown at cryogenic temperatures. Out of these effects, the electro-optic effect has great advantages for cryogenic applications since it only requires the generation of an electric field with electrodes and is preserved at cryogenic temperatures \cite{YOSHIDA1992,Morse1994b,Herzog2008,Thiele2020b}. We need to know how the electro-optic effect changes with a change in the temperature to integrate modulators with other optical components at cryogenic temperatures.

    A change in the refractive index $\Delta n_{ij}$ can be induced by an electric bias field $E_{DC}$ through the electro-optic effect with a given effective refractive index of the waveguide $\tilde{n}_i$ via \cite{Nye1957,Weis1985}
    \begin{equation}
            \Delta \left(\frac{1}{n_i^2}\right)=r_{ij} E_j^\mathrm{DC} 
     \end{equation}
     \begin{equation}
            \Delta n_{i} \approx -\frac{\tilde{n_i}^3}{2}r_{ij} E_j^\mathrm{DC}.
    \end{equation}
     The phase modulator uses this retardation to delay the propagation through the waveguide such that the output phase is shifted. Furthermore, the change of the refractive index can be used to couple two optical modes. The interaction of the optical modes is used to couple spatial modes in the directional coupler and polarisation modes in a polarisation converter. The resulting coupling between the optical modes is described by the coupled mode equations \cite{Yariv1973}. These equations can be reduced to a linear equation with the output state $A_{out}$ for a given input state $A_{in}$ using the transfer matrix $M$
     \begin{equation}
     	A_\mathrm{out}=M  \times A_\mathrm{in}.
    \end{equation}
    The input state is given by the input amplitude $a_{m,l}$ either for the spatial channels $a_1$ and $a_2$ or the polarisation states $a_H$ and $a_V$;
    \begin{equation}
       A=\left( a_{m},a_{l}\right)^{T}. 
    \end{equation} 
    
    The transfer matrix $M$ summarises the conversion from the input to the output depending on the coupling parameter $\kappa$;
    \begin{equation}
        \kappa =\frac{\pi \tilde{n}_i^3 r_{ij}  V \eta}{\lambda G}.
        \label{eq:kappa}
    \end{equation}
    The coupling parameter scales the interaction between the induced electric field $E^\mathrm{DC}$ and the supported optical modes $E_{m,l}$ in the waveguide. The overlap $\eta$ between the applied electric field and spatial modes is given by the overlap integral with a given electrode separation $G$ and applied voltage $V$;
    \begin{equation}
        \eta = \frac{G}{V} \frac{\int_A E^{*}_{m}  E^\textrm{DC} E_{l}\mathrm{d} A}{\sqrt{ \int_A E^*_{m} E_{m} \mathrm{d} A \int_A E^*_{l} E_{l} \mathrm{d} A}}.
        \label{eq:eta}
    \end{equation}
    Electro-optic modulation is additionally scaled by the material parameter $r_{ij}$. We expect a temperature-dependent change of the coupling since the electro-optic coefficient $r_{ij}$ and the refractive index $\tilde{n_l}$ are temperature dependent \cite{Herzog2008,Jundt1997a}.
    
    \subsection{Cryogenic modulation} \label{sec:eom}
        The coupling parameter $\kappa$ links the applied electric field and the output state of the modulator. For each modulator, the quantity $\Phi=\kappa L$ can be defined. For the phase modulator,  $\Phi$ is the phase induced given a bias voltage; $V_\pi/2$ is the voltage required to shift the phase by $\pi/2$. For the directional coupler, $\Phi$ describes the "phase" of the power transmission between the two output ports: a change of $\Phi = \pi/2$ is required to modulate the transmission from one output channel to the other. Similarly for the polarisation converter, $\Phi = \pi/2$ is required to transfer power fully from one polarisation mode to the other.
        
        

        The required voltage for the $\pi/2$-phase shift in each case is determined as the voltage $V_{\pi/2}$. A temperature-dependent decrease of the electro-optic coefficient $r_{ij}$ leads to an increase in the required modulation voltage $V_{\pi/2}$. 
        
       The electro-optic tensor element $r_{ij}$ acts with a different magnitude in different orientations of the induced electric field \cite{Nye1957,Weis1985}.  Different modulator types use different $r_{ij}$ for the modulation in our z-cut lithium niobate waveguides. The induced electric field uses for the modulation for the horizontal polarisation with $r_{13}$, for the vertical direction with $r_{33}$  and  for the coupling from the horizontal to vertical polarisation with $r_{24}$ \cite{Alferness1982,Sharapova2017a}. The polarisation modulator and the directional coupler use the electro-optic coefficient depending on the incident polarisation. Additionally, the polarisation converter uses the coefficient $r_{24}$ to convert the incident polarisation to the perpendicular polarisation. The electro-optic effect is temperature dependent, and decreases with the decrease of temperature \cite{PropLN}. It is to be determined if the electro-optic coefficient experiences the same change for all three orientations when cooled down from ambient to cryogenic temperatures.

        In addition to the temperature change, the coupling parameter $\kappa$ is inversely proportional to the wavelength $\lambda$ of operation. To compare a modulator at different operation wavelengths, the operation voltage $V_{\pi/2}$ should be scaled to the operation wavelength. 
        
        The thermal contraction is approximately 0.1\%  in lithium niobate from ambient temperatures to cryogenic temperature \cite{PropLN}, therefore, the added change of the output phase $\Phi$ due to thermal contraction is negligible with the decrease of temperature.

    \subsection{Temperature-dependent dispersion}
        
        All devices depend on modifying the refractive index with an external electric field. Due to dispersion, this refractive index is wavelength and temperature dependent. Understanding this interplay is crucial to build devices that function at a given wavelength under the same operating conditions.
        
\begin{figure}[H]
    		\centering
    		\includegraphics[width=0.5\linewidth]{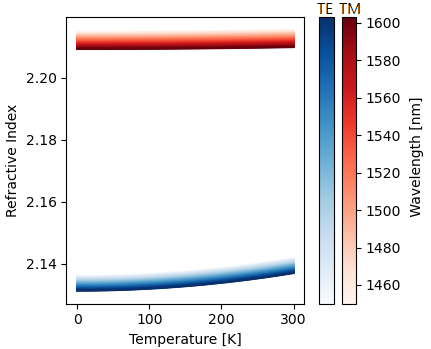}
    		\caption{Temperature dependent dispersion of the waveguide. The refractive index is extracted from the Sellmeier equations and the simulated titanium diffusion profile is for our  in-diffused waveguides.}
    		\label{fig:ReplotRefractiveIndex}
    	\end{figure}

        The refractive index of lithium niobate is described by the Sellmeier Equations \cite{Jundt1997a}. We simulated the temperature-dependent dispersion for our waveguide geometry, see Fig.~\ref{fig:ReplotRefractiveIndex}. To do so, the refractive index profile of the in-diffused titanium is simulated in dependence  of the temperature and wavelength with the simulation program RSoft \cite{Synopsys2021}. The refractive index for the vertical polarization TM remains almost unchanged by 0.03\% while the refractive index TE changes by 0.3\% in a range from 300K to 1K at 1550$\mathrm{nm}$.
        
\section{Fabrication}\label{sec:Fab}
    The modulators are fabricated from congruently grown z-cut lithium niobate wafers and the waveguides are realised by titanium in-diffusion \cite{Sohler2008c}. To do so, a layer of titanium is deposited on the samples top surface. Afterwards, a photoresist is deposited and selectively cured with UV mask lithography or UV laser lithography to structure the sample. Then, the titanium is etched such that a titanium stripe results on the surface along the entire sample length with the values given in figure \ref{tab:char}. To raise the refractive index in the sample the titanium stripe is in-diffused. The endfaces of the samples are polished to enable efficient optical coupling. The resulting waveguides achieve losses below 0.15$\,\mathrm{dB/cm}$ at 1550$\,\mathrm{nm}$. 
    
    Modulation in the waveguides is induced by an electric field originating from electrodes on the top surface. To fabricate and align the electrodes, a photoresist is structured with UV-lithography. The electrode pairs are separated by a gap $G$ over the full length $L$ of the electrodes with a electrode width $w$, as given in table \ref{tab:fab}. The direction of the electric field in the waveguide can be chosen by the placement of the electrode either in the vertical direction when placed on the top of the waveguide or in the horizontal direction when placed adjacent to the waveguide \cite{Kim1989}. The phase modulator has an electrode placed on the top of the waveguide to generate mainly the electric field vertically in the waveguide and an adjacent ground electrode. The electrodes are placed on top of the waveguides of the coupling region for the directional coupler. The polarisation converter requires a horizontal electric field. Therefore the electrodes are placed symmetrically next to the waveguide. Afterwards the electrode material is sputtered with a buffer layer of SiO$_2$ to reduce waveguide losses, a Cr layer for adhesion and an Au layer for the electrodes, with the given values in table \ref{tab:fab}. A lift-off process removes the unwanted electrode material  and the electrodes are contacted with wirebonds on the top. The chosen fabrication parameters of the modulators are optimised for the operation at ambient temperatures and summarised in table \ref{tab:fab}. The cryogenic operation of these modulators will need additional optimisations in the future since the overlap between the waveguide mode and electric field is likely to change due a temperature dependence in the refractive index and permittivity \cite{Jundt1997a,Herzog2008}.
    
    The polarisation converter requires an additional periodic poling for phasematching, as described in Sec.~\ref{sec:PolCon}. The waveguides are poled prior to electrode deposition. To do so, a photoresist with the required poling period is structured and liquid electrodes are deposited on the sample surface for biasing. RF voltage pulses invert the material domain under the electrodes, resulting in the required poling period. For our polarisation modulator this period is 19.9~$\upmu$m.

	\begin{table}[h!]
        \centering
        \begin{tabu}{|c|[2pt]c|c|[2pt]c|c|c|[2pt]c|c|c|}
        \hline
              Type&\multicolumn{2}{c|[2pt]}{waveguide}&\multicolumn{3}{c|[2pt]}{electrode}&\multicolumn{3}{c|}{electrode height}\\\hline
                  &Ti width&length&length&gap&width&SiO$_2$&Cr&Au\\\hline
             unit   &$\upmu$m&mm&mm&mm&$\upmu$m&nm&nm&nm                                                \\\tabucline[2pt]{-}
             Phase modulator       &5&22&12&9&100&400&10&300\\\hline
             Directional coupler   &7&27&12&6&100&400&10&100\\\hline
             Polarisation converter&6&22&15&15&400&400&10&100\\\hline
        \end{tabu}
        \caption{Summary of the fabrication parameters for the realised modulators.}
        \label{tab:fab}
    \end{table}

\section{Electro-Optical Characterisation}\label{sec:Char}
    We want to determine the voltage, wavelength, and temperature dependence of the three modulators. For the directional coupler and the polarisation modulator we measure the optical intensity in the two different modes while we sweep the bias voltage. These sweeps are done at different wavelengths of the input laser, and the wavelength sweeps are acquired at ambient and cryogenic temperatures. The phase modulator is characterised with the Sénarmont method \cite{Aillerie2000}, where we sweep the voltage at constant wavelengths. This is again done at different wavelengths and at ambient and cryogenic temperatures.
    
    
    To characterise the modulators, they are placed in a cryostat with windows for in- and outcoupling, as it can be seen in figure \ref{fig:phase} a), figure \ref{fig:koppler} a), and figure \ref{fig:polconmap} a). The light propagates through the sample and is modulated by applying a voltage across the electrodes. For the analysis of the output intensity the light is split by their spatial or polarization modes. The output of the phase modulator and polarisation converter are split with a polarising beam splitter, as it can be seen in figure \ref{fig:phase} a) and \ref{fig:polconmap} a). The output of the directional coupler is collimated and then split spatially with a D-shaped mirror, as it can be seen in figure \ref{fig:koppler} a). The resulting output intensity is then acquired by photodiodes.
    
    The voltage is swept in a range from -60 to 60V. We use a tunable continuous wave laser with a tuning range from 1440$\,\mathrm{nm}$ to 1640$\,\mathrm{nm}$, a linewidth of 1$\,\mathrm{pm}$, and an intensity of 1$\,\mathrm{mW}$. To avoid changes in the setup, the sample is mounted in the cryostat for the characterisation at both ambient and cryogenic temperatures, as it is summarised in table \ref{tab:char}. 
    
    \begin{table}[h!]
        \centering
        \begin{tabu}{|c|[2pt]c|c|[2pt]c|c|[2pt]c|c|}
        \hline
             Type&\multicolumn{2}{c|[2pt]}{wavelength [nm]}&\multicolumn{2}{c|[2pt]}{voltage [V]}&\multicolumn{2}{c|}{temperature [K]} \\\hline
                                   &range&steps&range&steps&max&min     \\\tabucline[2pt]{-}
             Phase modulator       &1440-1640&2.5&$\pm$60&1&296&8.5\\\hline
             Directional coupler   &1440-1640&2&$\pm$55&0.5&296&5\\\hline
             \multirow{2}{*}{Polarisation converter}&1464-1480&0.2&$\pm$60&1&296&\\
                                   &1578-1590&0.2&$\pm$60&1& &5\\\hline
        \end{tabu}
        \caption{Parameter range of the wavelength and voltage sweeps. The parameters sweeps are given for all three modulators at ambient and cryogenic temperatures.}
        \label{tab:char}
    \end{table}

\section{Modulators}\label{sec:Modulators}

    \subsection{Phase Modulator}
        A phase modulator demonstrates the core principle of  electro-optic modulation. The modulator changes the refractive index in the waveguide and a tunable propagation delay is induced. This delay can be used to modulate the interference in an integrated Mach-Zehnder-interferometers or the phase between path-entangled states in quantum optics. Electro-optic phase modulators are therefore an essential tool for quantum optic networks \cite{Sharapova2017a,Luo2019}. Additionally, a lithium niobate phase modulator has been used to read out superconducting devices at cryogenic temperatures \cite{Youssefi2021}. 
               
        An electric field induces the change in the refractive index via the electro-optic effect depending on the incident polarisations, as described in section \ref{sec:eom}. The accumulated phase $\Phi_{(H,V)}$ after the propagation through a waveguide can be described by the transfer matrix for the input in both polarisation modes written in vector form $A_{in}=(a_H,a_V)^T$:
    	\begin{equation}
  		    M_\mathrm{Phase}=\begin{pmatrix} \exp(i\Phi_{H})& 0\\ 0 & \exp(i\Phi_{V}) \end{pmatrix}.
    	\end{equation}
        The phase at the output of the waveguide due to the electro-optic modulation is then $\Phi_\mathrm{EO}=\kappa_{(H,V)} L$, which is described in section \ref{sec:theory}. The coupling parameter is given by the equation \ref{eq:kappa} depending on the polarisation with $r_{13}$ and $r_{33}$ respectively.

        The Sénarmont method can read out the phase difference of two polarisation modes by interfering them after the transmission through the sample \cite{Parravicini2011}. Light is coupled equally in the waveguide and modulated by the electric field with an accumulated phase difference
        \begin{equation}
        	\Delta \Phi = \kappa_{H} L -\kappa_{V} L.
        \end{equation}
        The output of both polarisations are then interfered with an quarter waveplate at 45$^\circ$ and a Polarising Beam Splitter (PBS). The intensity $I$ in one arm after the PBS is then described by:
        \begin{equation}
       	 I_{1,2}=\frac{1}{2}(1\pm\sin(\Delta \Phi)).
        \end{equation}

The electric field generated by the electrode on the top of the waveguide mainly modulates the phase in the vertical direction and partially in the horizontal direction, due to the horizontal geometry of the electrodes. 
 
\begin{figure}[H]
			\centering
			\includegraphics[width=0.8\linewidth]{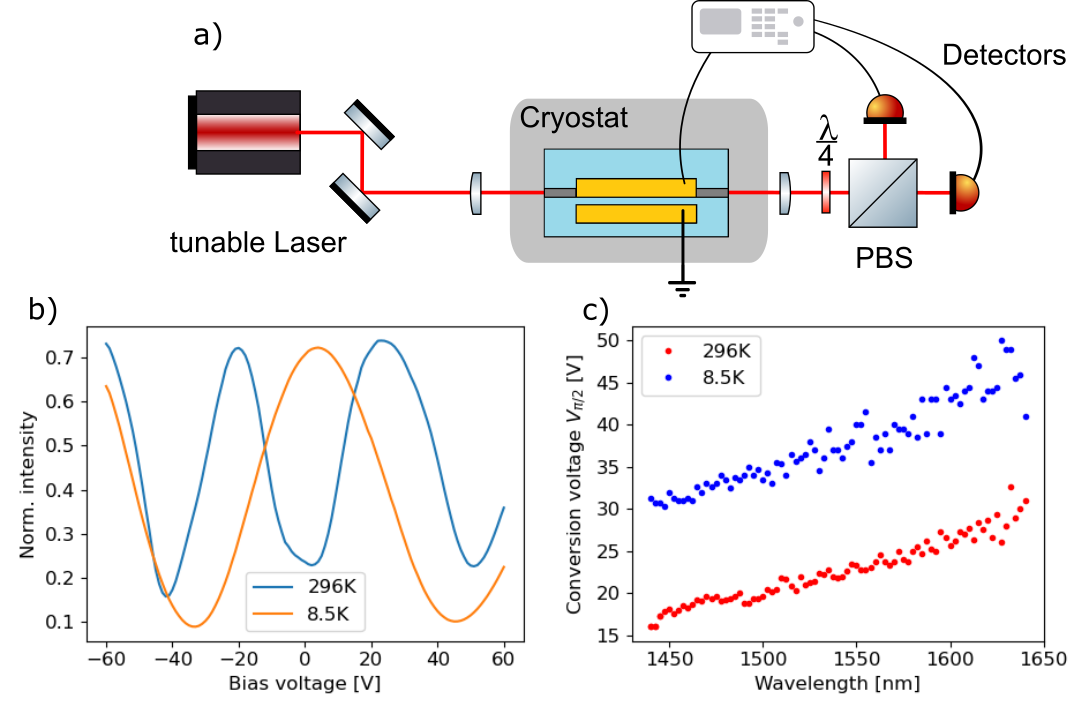}
			\caption{a) The electro-optic phase shifter is integrated in a cryostat for optical testing. A tunable laser is coupled in the waveguide. The output characterised with a quarter wave plate and a polarisaing bema splitter (PBS). b) Sweep of the bias bias voltage at 1550$\,\mathrm{nm}$. c) Wavelength dependent $V_{\pi/2}$}
			\label{fig:phase}
	   	\end{figure}

A single voltage scan at 1550$\,\mathrm{nm}$ is shown in figure \ref{fig:phase} b). We normalised the output of the modulator by the sum of the intensity in both channels of the PBS: $I_{norm}= I_{H,V}/( I_{H}+ I_{V})$. The resulting modulation gives the expected $sine$-shape. The modulation voltage $V_{\pi/2}$ is the voltage difference required to modulate the output intensity from a minimum to a maximum. Propagation and coupling losses reduce the intensity of the beams and therefore the visibility of the modulation. Nonetheless, the  $V_{\pi/2}$-voltage is unaffected from the induced losses. 
 
 The modulation voltage $V_{\pi/2}$ increased from 23.3$\,\mathrm{V}$ at ambient temperatures to 40$\,\mathrm{V}$ at cryogenic temperatures. In addition, we extracted the  $V_{\pi/2}$ from the wavelength scan at different temperatures as it can be seen in figure \ref{fig:phase} c). A linear increase of the modulation voltage is observed with the increase of the operation wavelength. A DC-drift in the modulation characteristics was not observed during the characterisation of the modulator over the duration of a day. 
 
 The phase modulator was operated successfully at ambient temperatures and at cryogenic temperatures over the full wavelength range from 1440-1640$\,\mathrm{nm}$. The modulation voltage $V_{\pi/2}$ increased with the decrease in temperature and increase of the operation wavelength. In this case, the operating temperature was 8.5K, due to increased electrical heat load in relation to the other tested modulators. We used the Sénarmont method to characterize the phase modulator because this method shows the core principle of the electro-optic modulation. Additionally, the method allows us to use only a single input beam. The phase modulator can be used on a single polarisation but a stable reference beam must be used to read out the phase. The resulting modulation voltage will be lower because the phase difference of only one polarisation is read out. Nonetheless, this method for the characterisation shows the principle of the electro-optic modulation at cryogenic temperatures.

    \subsection{Directional coupler}
        Electro-optic directional couplers are used in integrated photonics to route light between waveguides. This effect is used for the switching of single photons \cite{Luo2019} or in macroscopic applications for intensity modulation \cite{Schmidt2004}. 
        
\begin{figure}[H]
			\centering
			\includegraphics[width=0.8\linewidth]{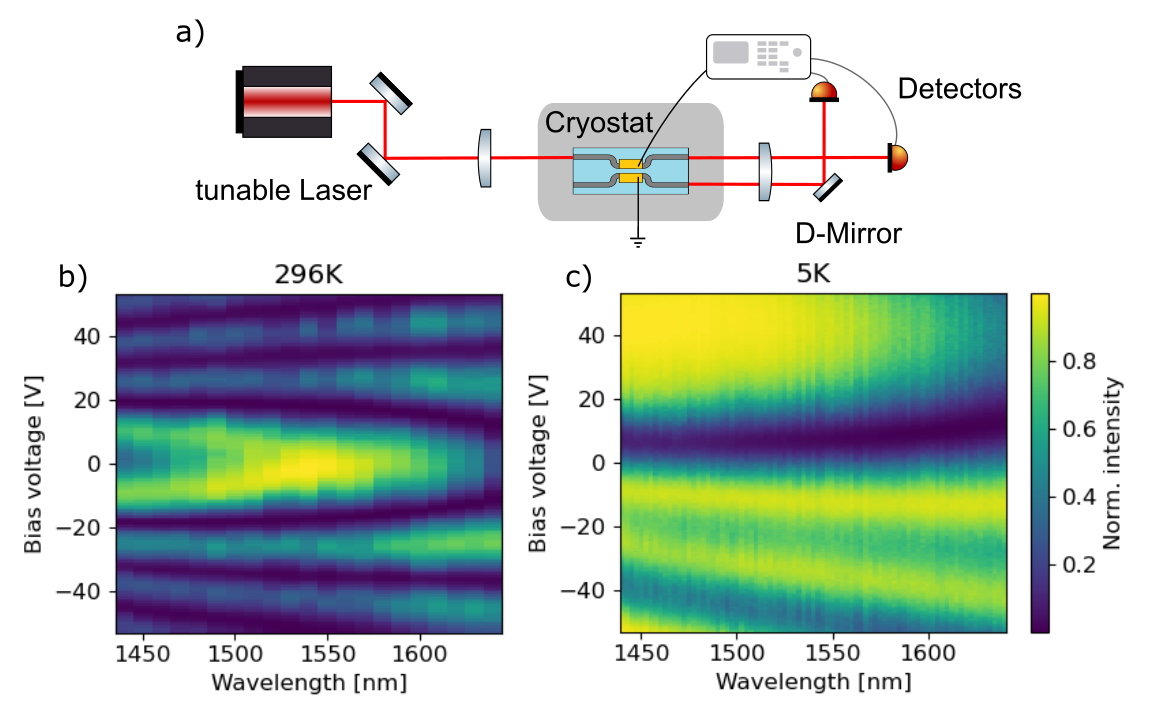}
			\caption{ a) The coupler is integrated in a cryostat for optical characterisation with a polarized tunable laser and a bias voltage supply. The resulting 2D plot of the wavelength dependent electronic modulation is given at b) room temperature and c) at 5\,K. The intensity in the 2D plot is normalized by the sum of both intensities; $I_{1}/(I_{1}+I_{2})$}
			\label{fig:koppler}
	   	\end{figure}

        Light is guided through a waveguide towards an adjacent waveguide such that the evanescent field of both guided modes overlap. After the interaction, bends divert the waveguides and guide the coupled light apart, as seen in figure~\ref{fig:koppler} a). The output of the coupler depends on the propagation in the waveguides, which for symmetric waveguides is described by the propagation constant $k$.
        
        The electrodes on the top of the waveguides generates a vertical electric field and induces a phase difference:
        \begin{equation}
            \beta = \kappa_1-\kappa_2 = \frac{\pi \tilde{n}_i^3 r_{ij} \eta (+V)}{\lambda G}- \frac{\pi \tilde{n}_i^3 r_{ij}  \eta (-V)}{\lambda G}~.
        \end{equation}
        Due to the electrode configuration the electric field in both waveguides are orientated in opposing directions. The electro-optic modulation is therefore dependent on the overlap between the spatial mode of the adjacent waveguides $E_{1,2}$ and the induced electric field $E^{DC}$ \cite{Kim1989}     
        \begin{equation}
            \eta = 2 \frac{G}{V} \frac{\int_A E^{*}_{1}  E^{DC} E_{2}\mathrm{d} A}{ \int_A E^*_{1} E_{2} \mathrm{d} A}
        \end{equation}
        The state of the modulator at a given propagation distance $y$ is described by the transfer matrix for an input state $A_{in}=(a_1,a_2)^T$ for both waveguides \cite{Kogelnik1976a}
          \begin{equation}
    	    M_\mathrm{directional\, coupler}=
    		\begin{pmatrix} \cos(sy)+ \frac{i \beta}{2 s} \sin(sy)	 & -  \frac{k}{s} \sin(sy)\\ +  \frac{k}{s} \sin(sy) & \cos(sy)- \frac{i\beta}{2 s} \sin(sy) \end{pmatrix}
            \end{equation}
          The output state of the directional coupler at $y=L$ is dependent on the switching parameter $s=\sqrt{k^2+(\beta/2)^2}$. The induced electric field thus changes the propagation in the coupler by changing the switching parameter $s$.

        The directional coupler requires a voltage of 18V to couple TM-polarised light from the input waveguide to the adjacent waveguide at 1550nm and ambient temperatures. The $V_{\pi/2}$-voltage increases to about 35V at 1550nm and cryogenic temperatures. The wavelength-dependent output of the modulator at ambient temperatures resembles the wavelength dependent modulation with a shift of 42V at cryogenic temperatures. This DC-shift could be induced by localised charges in the sample generated during the cooling process due to the pyro-electric effect \cite{Nye1957,PropLN}. The DC-shift remained stationary during the characterisation of the device over the course of more than a day.   

        Additionally, the comparison of the wavelength-dependent modulation can be determined by comparing the output when the light remains in the same waveguide, known as the bar state. This bar state shifted from 1545$\,\mathrm{nm}$ at ambient temperatures to 1460$\,\mathrm{nm}$ at cryogenic temperatures, a shift of 85 $\,\mathrm{nm}$. The output of the directional coupler is wavelength-dependent since dispersion changes the coupling ratio of the waveguide modes.
        
    \subsection{Polarisation Converter}\label{sec:PolCon}
        Controlling the polarisation on a single photon level is of great interest for quantum optics. Manipulating photonic qubits can be realised by polarisation conversion in combination with frequency conversion \cite{Luo2019}.
        
        The modulator converts incident light from one polarisation to the orthogonal polarisation using the electro-optic effect. The conversion is inefficient at a given wavelength due to differences in the propagation speeds of the two polarisation modes, since z-cut lithium niobate waveguides are birefringent. The mismatch in propagation speed results in an acquired phase difference $\Delta \beta$.  A periodic poling of the crystal axis along the propagation direction can compensate the accumulated phase difference \cite{Yariv1973,Thiele2020b}. Therefore,  a poling length $\Lambda$ should be chosen such that
        \begin{equation}
        	\Delta \beta = 2 \pi \left(\frac{n_\textrm{H}(\lambda,T)-n_\textrm{V}(\lambda,T)}{\lambda}\right)-\left(\frac{2 \pi}{\Lambda}\right) = 0.
        \end{equation}
    With a designed periodic poling  for a given operation wavelength the polarisations couple efficiently. The transfer matrix that describes the polarisation conversion from $H$ to $V$ when operated on a single waveguide with the input $A_{in}=(a_H,a_V)^T$ is given by     	
    \begin{equation}
    	    M_\mathrm{pol\, conv}=
    		\begin{pmatrix} \cos(sy)+ \frac{i\Delta \beta}{2 s} \sin(sy)	 & -  \frac{\kappa}{s} \sin(sy)\\ +  \frac{\kappa}{s} \sin(sy) & \cos(sy)- \frac{i\Delta \beta}{2 s} \sin(sy) \end{pmatrix} \cdot e^{\frac{i \Delta \beta y}{2}}~.
            \end{equation}
The coupling strength between the polarisation modes is dependent on the switching parameter $s=\sqrt{\kappa^2+\left({\Delta \beta}/{2}\right)^2}$. Coupling is most efficient at the phasematched wavelength with $\Delta\beta =0$, such that the switching parameter is equal to electro-optic coupling $s=\kappa$. The coupling parameter $\kappa$ is given by equation \ref{eq:kappa} with the effective refractive index as the geometric mean $\tilde{n}=(n_{H}n_{V})^{1/2}$ and the electro-optic coefficient $r_{24}$.
        
		\begin{figure}[H]
			\centering
			\includegraphics[width=0.80\linewidth]{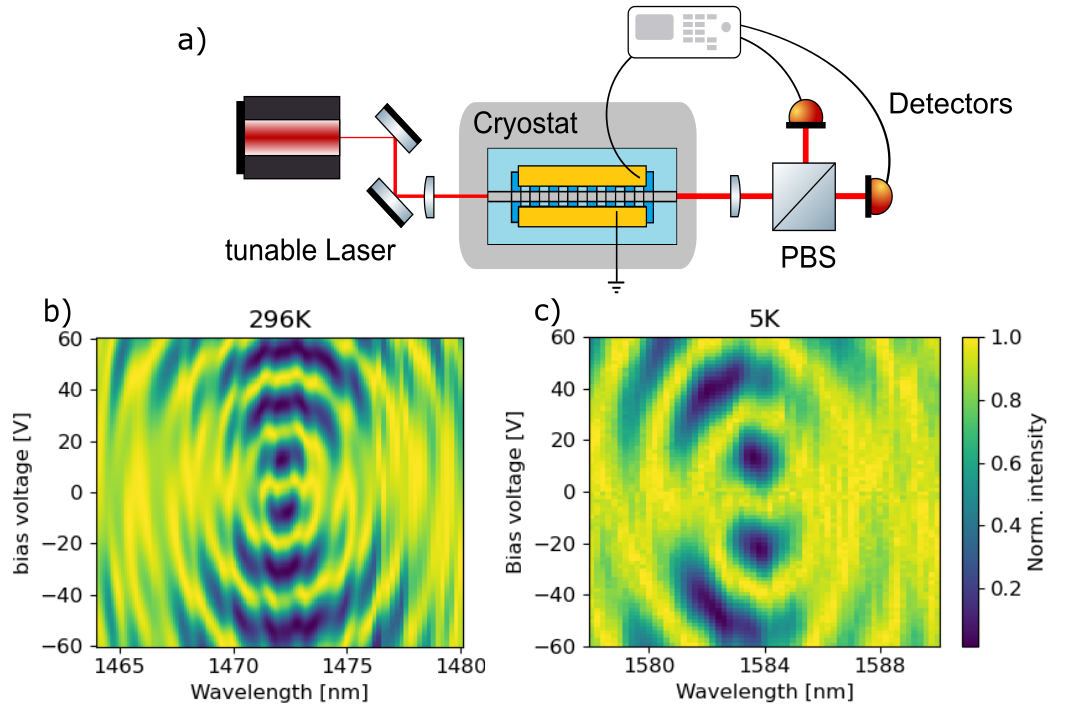}
						\caption{ a) The coupler is integrated in a cryostat for optical characterisation with an tunable input laser and a bias voltage supply. The resulting 2D-plot of the wavelength dependent electronic modulation is given at room temperature b) and at 5\,K c). The intensity is in the plot is normalized by the sum of both intensities. $I_{H,V}/(I_{H}+I_{V})$}
			\label{fig:polconmap}
	   	\end{figure}

         The change of the operation temperatures affects the refractive index of the waveguides such that the phasematched wavelength shifts. The results of the characterisation at 296$\,\mathrm{K}$ and 5$\,\mathrm{K}$ can be seen in figure~\ref{fig:polconmap} b) and c). 
        
        The phasematched wavelength of the converter shifted from 1472$\,\mathrm{nm}$ to 1584$\,\mathrm{nm}$ for a poling period of 19.9\,$\upmu$m from 295$\,\mathrm{K}$ to 5$\,\mathrm{K}$. The required conversion voltage $V_{\pi/2}$, at the phasematched wavelength, increased from 11.1$\,\mathrm{V}$ to 16.1$\,\mathrm{V}$. The increase in the required conversion voltage is linked to the decrease of the electro-optic coefficient $r_{24}$. In addition, a DC-offset is present in the range of 2.1$\,\mathrm{V}$ and remained stationary during the characterisation over more then one day.
        
        The modulation map at cryogenic temperatures resembles closely the shape of the conversion map at ambient temperatures. A wavelength shift of 112$\,\mathrm{nm}$, an increase of $V_{\pi/2}$ of 5\,$\mathrm{V}$, and a DC-shift of 2.1V is observed with the decrease of temperature.  Additionally, we observed a reduced perturbation of the conversion characteristics in comparison to a previous cryogenic characterisation of a polarisation converter \cite{Thiele2020b}.

\section{Device comparison}\label{sec:Comp}
 In summary, we realised three different devices to characterise the electro-optic modulation at cryogenic temperatures. First, a phase modulator inducing a phase difference was characterised with the Sénarmont method. Second, a directional coupler was realised to modulate the routing from one waveguide to the other. Lastly, a polarisation converter was realised to couple light from an incident polarisation to the orthogonal polarisation. This modulator toolbox can be used to realise integrated photonic circuits at cryogenic temperatures. We now compare the performance of these modulators at ambient temperatures and cryogenic temperatures.
 
The required modulation voltage $V_{\pi/2}$ and operation wavelength are summarised in the table \ref{tab:summary}. All modulators showed an increase of the operation voltage $V_{\pi/2}$ at cryogenic temperatures, namely 74\% for the phase modulator, 84\% for the directional coupler and 35\% for the polarisation converter, as shown in table \ref{tab:summary}. Note that the effect of different operating wavelengths and electrode lengths are normalised out in each case. The increase in the modulation voltage can be attributed to the material parameters connected to the modulation strength $\kappa$ such as the electro-optic coefficient $r_{ij}$ and the mode overlap $\eta$. For example, previously the electro-optic coefficient $r_{33}$ has been reported to decrease by ca. 20\% for bulk lithium niobate at cryogenic temperatures \cite{Herzog2008}. This decrease would be equivalent to a voltage increase of 28\% for the directional coupler used in this work. The discrepancy with the 85\% increase in the $V_{\pi/2}$-voltage which we observe could be due to a reduced overlap between the guided mode and the electric-field at cryogenic temperatures. The overlap can be further optimised by the electrode placement and dimension at cryogenic temperatures. The other modulators use different $r_{ij}$ parameters, however changes to the mode overlap make it difficult to infer changes in this parameters under cryogenic conditions.

     \begin{table}[h!]
        \centering
        \begin{tabu}{|c|[2pt]c|c|[2pt]c|c|[2pt]c|c|}
        \hline
             &\multicolumn{2}{c|[2pt]}{$V_{\pi/2}$ [$V$]}   &\multicolumn{2}{c|[2pt]}{$\lambda [nm]$ } &\multicolumn{2}{c|}{$V_{\pi/2} /(\lambda L)$  [$\frac{V}{\upmu m\,cm}$]}\\\hline
             Type& 296K&$<$8.5K&296K&$<$ 8.5K&296K& $<$ 8.5K\\\tabucline[2pt]{-}
             Phase modulator&23.3& 40 &\multicolumn{2}{c|[2pt]}{1440\,-\,1640 }&12.4&21.5\\\hline
             Directional coupler &18& 35&1460&1545&10.2&18.8\\\hline
             Polarisation converter &11.1& 16.1&1472&1584&5&6.73\\\hline
        \end{tabu}
        \caption{Summary of the results from the temperature dependent modulator characterisation. The given wavelengths are the bar state wavelength for the directional coupler and phase matched wavelength for the polarisation converter.}
        \label{tab:summary}
    \end{table}
 
The operation of the directional coupler and polarisation converter showed a DC-shift in the modulation characteristics at cryogenic temperatures. We associate this DC-shift to be generated by pyro-electric charges during the cooling process \cite{Nye1957,PropLN}. These charges are localised in the substrate such that a static electric field is generated. During the cooling process the charges can be discharged and generated, scaled by the charge mobility of the substrate. The charge mobility in lithium niobate decreases at lower temperatures such that the charges are less likely to discharge \cite{PropLN}. The resulting electric field builds up and presents as a DC-shift in the modulation at cryogenic temperatures. This DC-shift remained stable under cryogenic conditions since the charge mobility in the substrate is greatly reduced and additional DC-drifts could be less likely.

The output of the modulators shows a wavelength dependence. To match multiple optical devices the interaction length can be further optimised to achieve the desired output state at cryogenic temperatures. The temperature induced wavelength shift in the unbiased bar state is 84$\,\mathrm{nm}$ for the directional coupler and 112$\,\mathrm{nm}$ for the polarisation converter.  The desired operation wavelength can be optimised by either optimising the coupler length and waveguide separation or by optimising the periodic poling. The phase modulator can be operated in the entire wavelength range. Additionally, a wavelength-dependent $V_{\pi/2}$ was determined. The modulation voltage increase is expected to be 14\% given the relation: $\kappa \approx 1/\lambda$. However an additional increase by 31\% is observed at ambient temperatures and at cryogenic temperatures. The wavelength dependent increase could be linked to a reduced overlap between the optical mode and induced electric field.

\section{Conclusion}\label{sec:concl}
We successfully realised a toolbox for light manipulation for integrated optics at cryogenic temperatures. The modulators are fabricated with titanium in-diffused z-cut lithium niobate waveguides. We fabricated and characterised a phase modulator, directional coupler and polarisation converter and investigated changes in the electro-optic effect at cryogenic temperatures. The modulation voltage $V_{\pi/2}$ increases with the decrease of temperature for all three modulators, and can be optimised further for these devices by optimising the overlap between the optical modes and induced electric field. Future characterisation can focus on a reduction of pyro-electricity during the cooling process. 

The future goal for the cryogenic modulators is the integration with SNSPDs. Integrated SNSPDs have been shown before with system detection efficiencies of about 1\% for in-diffused waveguides and thin lithium niobate at 46 \% \cite{Tanner2012,Hopker2019,Lomonte2020,Lomonte2021}. The evanescent coupling from the waveguides to the detectors may need further improvements depending on the desired application. Additionally, feed-forward applications could be realised by the combined integration. An impinging on the SNSPD generates an electronic signal which modulates light in the integrated electro-optic modulator. The amplitude of the detection signals is in the range of a few millivolts while the modulation voltage $V_{\pi/2}$ is above 16$\,\mathrm{V}$ at cryogenic temperatures. Additional amplifiers are then needed to generate a strong optical modulation based on a single photon click. This characterisation of the electro-optic modulator is a step forwards towards the combined operation of integrated detectors and modulators at cryogenic temperatures. 

\section{References}

\bibliographystyle{ieeetr}
\bibliography{main.bib}

\appendix
\end{document}